\documentclass[twocolumn,showpacs]{revtex4}
\usepackage{graphicx}
\usepackage{subfigure}
\usepackage{amsmath,amssymb,amsfonts}

\newcommand{\no}[1]{}
\begin{document}

\author{
E.~V.~Votyakov{$^{1,2\P}$}, Yu.~Kolesnikov{$^1$},
O.~Andreev{$^{1}$}, E.~Zienicke{$^{2}$}, A.~Thess{$^{1}$}}


\affiliation{
     $^1$Fakult\"at Maschinenbau, Technische Universit\"at Ilmenau,\\
$^2$Institut f\"ur Physik, Technische Universit\"at Ilmenau, PF
100565, 98684 Ilmenau, Germany}

\title{\sffamily Structure of the Wake of a Magnetic Obstacle}


\begin{abstract}
We use a combination of numerical simulations and experiments to
elucidate the structure of the flow of an electrically conducting
fluid past a localized magnetic field, called magnetic obstacle.
We demonstrate that the stationary flow pattern is considerably
more complex than in the wake behind an ordinary body. The steady
flow is shown to undergo two bifurcations (rather than one) and to
involve up to six (rather than just two) vortices. We find that
the first bifurcation leads to the formation of a pair of vortices
within the region of magnetic field that we call \textit{inner
magnetic vortices}, whereas a second bifurcation gives rise to a
pair of \textit{attached vortices} that are linked to the inner
vortices by \textit{connecting vortices}.
\end{abstract}

\maketitle

\begin{figure}[]
    \includegraphics[width=8.25cm, angle=0, clip]{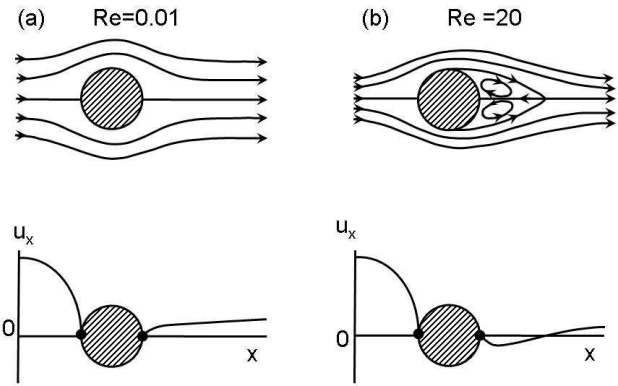}
    \caption{\label{Fig:Feynman}Streamlines of the steady flow around a solid
    cylinder and corresponding streamwise
    velocity along the centerline: (a) without vortices
    and (b)  with attached vortices.  Notice, the region of attached vortices
    corresponds to negative streamwise velocity.
    }
\end{figure}

When a liquid metal moves relative to a localized magnetic field -
a magnetic obstacle - the induced eddy currents produce a Lorentz
force that creates vorticity. Whereas the flow pattern around a
mechanical obstacle, such as the cylinder shown in
Fig.~\ref{Fig:Feynman}, are well documented
\cite{Feynman:Lectures:Vol2:1964}, the structure of the wake of a
magnetic obstacle is poorly understood even in the seemingly
simple steady state. In the present Letter we demonstrate that the
first bifurcation in the flow past a magnetic obstacle leads to
the formation of a vortex-pair inside the magnetic field and that
the formation of attached vortices, the analog to
Fig.~\ref{Fig:Feynman}$b$, is the result of a second bifurcation
giving rise to a remarkably complex six-vortex pattern.

\begin{figure}[b]
    \includegraphics[width=7.5cm, angle=0, clip]{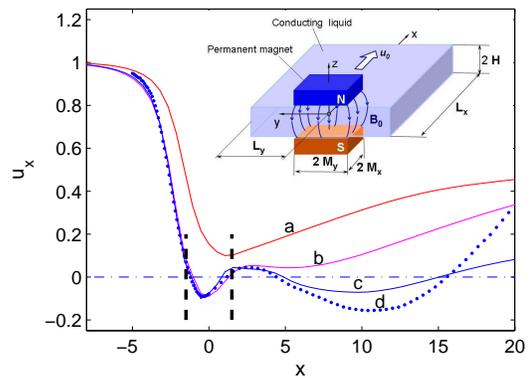}
    \caption{\label{Fig:Centerline}(color online) Magnetic obstacle (inset)
    and streamwise velocity along
    the centerline from 3D numerics($a-c$) and experiment ($d$): N=4, Re=100
    ($a$), N=11.25, Re=100 ($b$), N=11.25,
    Re=400 ($c$), N=11.25, Re=2000 ($d$).
    Dashed vertical lines mark borders of the magnetic obstacle.
    }
\end{figure}

\begin{figure*}[]
    \includegraphics[width=17cm, angle=0, clip]{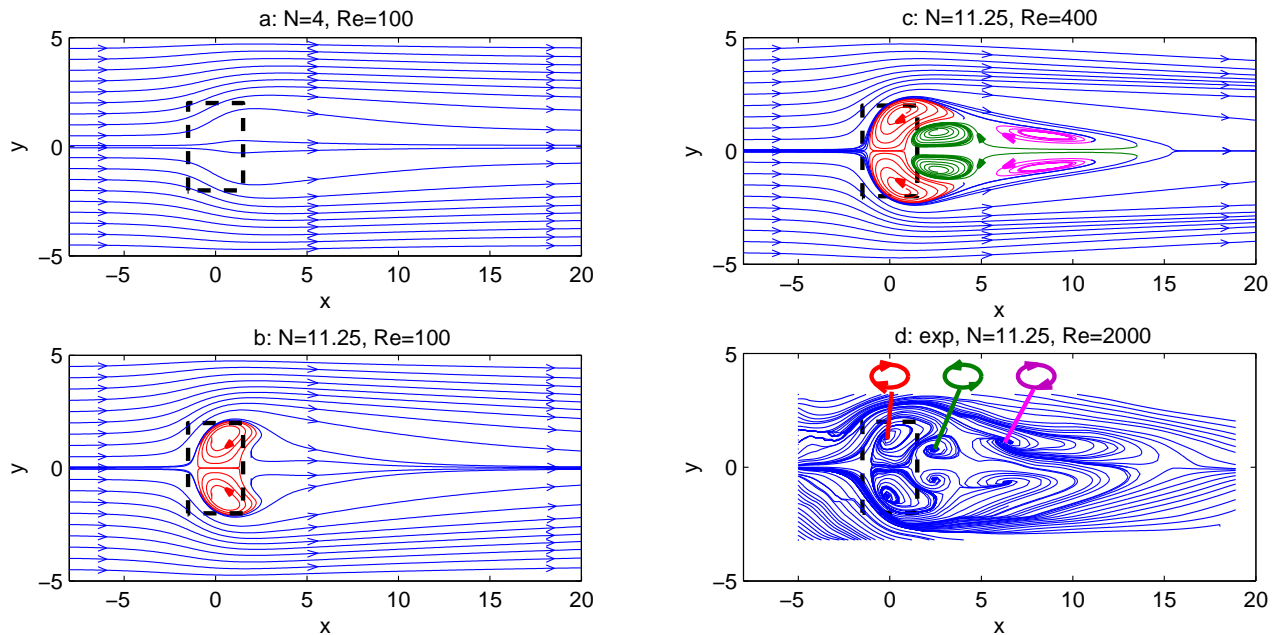}
    \caption{\label{Fig:2Dfields}(color online).
    Flow streamlines in the middle plane corresponding to profiles of
    Fig.~\ref{Fig:Centerline} involving inner (red $b-d$),
    connecting (green $c,d$), and attached (magenta $c,d$)
    vortices. Dashed lines mark borders of the magnetic obstacle. }
\end{figure*}

Understanding the wake of a magnetic obstacle, whose prototype is
shown in Fig.\ref{Fig:Centerline}, is of considerable fundamental
and practical interest. On the fundamental side
\cite{Shercliff:book:1962,Roberts:1967,Davidson:book:2001,Moreau:book:1990},
a magnetic obstacle represents a system with a rich variety of
dynamical states whose behavior is governed by two parameters, the
Reynolds number $Re=u_0H/\nu$ (where $u_0$ is velocity, $H$ is the
characteristic linear size, and $\nu$ is the kinematic viscosity
of the fluid) and the magnetic interaction parameter $N=\sigma
HB_0^2/\rho u_0$ (where $\sigma$ is the electric conductivity of
the fluid, $\rho$ its density and $B_0$ the magnitude of the
magnetic field) \cite{Shercliff:book:1962}. On the practical side,
spatially localized magnetic fields enjoy a variety of industrial
applications \cite{Davidson:Review:1999} exemplified by
electromagnetic stirring, electromagnetic brake, and non-contact
flow measurement \cite{Thess:Votyakov:Kolesnikov:2006}.

An analogy between the hydrodynamic obstacle and a braking
localized magnetic field has been supposed from the beginning of
Magnetohydrodynamics.
However, up to now the corresponding reverse flow behind a
magnetic obstacle has never been observed experimentally. The
earliest 2D 
calculations \cite{Gelfgat:Peterson:Sherbinin:1978} showed a kind
of recirculation, but the especially designed experiments
\cite{Gelfgat:Olshanskii:1978} failed to confirm it. Recent 2D
numerical papers of Cuevas \textit{et~al.} revived  the term
'magnetic obstacle' \footnote{one of the authors, Yu.K., used
"magnetic obstacle" as a working term in the seventies in Riga,
MHD center of the former USSR} and found a Karman vortex street
\cite{Cuevas:Smolentsev:Abdou:2006} for large $Re$, as well as a
vortex dipole in a creeping flow
\cite{Cuevas:Smolentsev:Abdou:PRE:2006}.


We shall derive logically, for the first time, new possible
stationary magnetohydrodynamic flow patterns and illustrate them
by both physical experiments and 3D numerical simulations.
Technical details to reproduce our quantitative results are given
to the end of the paper.

Let us recall that interaction parameter $N$ represents the ratio
of the Lorentz forces to the inertial forces and Reynolds number
$Re$ is the ratio of inertial and viscous forces. From this we
derive that (i) the higher $N$ the stronger is the braking Lorentz
force \textit{inside} the obstacle and (ii) the larger $Re$ the
more pronounced is the stagnant region \textit{behind} the
obstacle. Now we analyze all possible dependencies of streamwise
velocity along the centerline (centerline curves),
Fig.~\ref{Fig:Centerline}, analogously as it is shown in
Fig.~\ref{Fig:Feynman}. Figure \ref{Fig:2Dfields} illustrates
these centerline curves by flow patterns obtained from 3D
numerical simulation and physical experiment. It is obvious to
conclude that streamwise velocity inside the magnetic obstacle is
positive for small $N$ (curve $a$) and might be negative when $N$
is higher than a critical value $N_c^m$ (curves $b-d$). Past the
obstacle the flow structure is analogous to that of a solid
cylinder shown in Fig.~\ref{Fig:Feynman}: at low $Re$ the velocity
is positive (curves $a,b$), and it might be negative when $Re$ is
larger $Re_c(N)$ (curves $c,d$).

It is a signal for a reverse flow when a centerline curve becomes
negative. Therefore, we deduce that in the most developed case
there must be a pair of vortices inside the obstacle, which we
call \textit{inner magnetic} vortices
and show by red color in
Fig.~\ref{Fig:2Dfields}($b-d$), and a pair of hydrodynamical
\textit{attached} vortices past the obstacle marked by magenta
color in Fig.~\ref{Fig:2Dfields}($c,d$).

\begin{figure}[]
    \includegraphics[width=6.0cm, angle=0, clip]{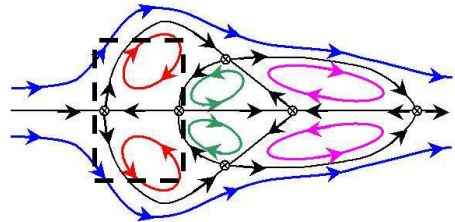}
    \caption{\label{Fig:Topology}(color online)
    Topology of the flow involving inner (red), connecting (green), and attached (magenta)
    vortices, bold dashed lines mark borders of magnetic obstacle. Cross-marked are fixpoints,
    black lines are heteroclinic orbits.
    }
\end{figure}

The inner vortices and the attached vortices show the same
direction of rotation determined by the main flow movement. This
means that a continuous flow only can exist, if the pairs of inner
and attached vortices are connected by contra-rotating
\textit{connecting} vortices shown by green color in
Fig.~\ref{Fig:2Dfields}($c,d$). These connecting vortices
correspond to a local maximum of centerline  curves in
Fig.~\ref{Fig:Centerline} at $x\approx 2.5$. The resulting flow
structure of Fig.~\ref{Fig:2Dfields}($c,d$) becomes clear from
Fig.~\ref{Fig:Topology}, where the 2D flow structure in the middle
plane is organized around six hyperbolic fixpoints  and their
corresponding heteroclinic invariant manifolds. As can be derived
from nonlinear topological dynamics
\cite{Guckenheimer:Holmes:1983} no other two-dimensional volume
conserving and continuous differentiable flow structure is
possible, in which two pairs of vortices with the same sense of
rotation can coexist. The inner and connecting vortices with the
surrounding invariant manifolds replace the rigid body of
Fig.~\ref{Fig:Feynman}. Notice, that a \emph{virtual} (magnetic)
obstacle induces \emph{real} vortices inside. This is a direct
representation of the hydrodynamic potential theory where a
\emph{real} physical body is modelled by means of a \emph{virtual}
vortex dipole flow.

Recirculation inside the obstacle is dependent on the interaction
parameter $N$ being inverse proportional to velocity. Therefore,
starting from a stable flow without vortices, inner vortices and
then the six-vortex pattern involving attached vortices appear by
\textit{decreasing} the flow rate. This is a paradoxical behavior
being in contrast to the ordinary hydrodynamics where attached
vortices always appear at \textit{increasing} flow rate.

To confirm the foresaid theoretical finding experiments have been
performed. The measured centerline velocity, curve $d$ in
Fig.~\ref{Fig:Centerline}, agrees very well with simulated results
in the region of the magnets. Behind the magnets there is a
discrepancy which is obviously explained by the fact that the
experimental $Re=2000$ is essentially larger than in 3D
simulation. However, the experimental flow structure shown in
Fig.~\ref{Fig:2Dfields}($d$) is qualitatively the same as in
Fig.~\ref{Fig:2Dfields}($c$). Therefore, the found six-vortex
pattern is stable in the experiments even at $Re=2000$. Both
experiment and numerical simulation show that with increasing $Re$
and constant $N$ the attached vortices grow while the connecting
vortices slightly shrink.


Direct quantitative comparison between experiment and simulation
is not possible due to the following reasons. First, inlet flow is
laminar in simulations and turbulent in experiments. Second, high
experimental Hartmann number 
is difficult to handle in a proper 3D simulation since Hartmann
layers must be finely resolved on the bottom and top walls. For
qualitative comparison these factors are not important, since $Re$
number plays no role for the core flow
\cite{Votyakov:Zienicke:FDMP:2006} and turbulent pulsations are
suppressed by magnetic field \cite{Andrejew:Kolesnikov:2006}. The
slight spanwise asymmetry of the experimental flow,
Fig.~\ref{Fig:2Dfields}$d$, is due to a bend of the flow circuit
in front of the channel inlet.

Let us now shortly describe some technical issues. The governing
equations which were solved in the present 3D simulation to obtain
Fig.~\ref{Fig:Centerline},~\ref{Fig:2Dfields} are composed of
Navier-Stokes plus Maxwell equations for moving medium and Ohm's
law. Induced magnetic field is assumed to be infinitely small
compared to an external magnetic field \cite{Roberts:1967}. Then,
these equations are the following:
    \begin{eqnarray*}
    \frac{{\partial\textbf{u}}}{{\partial t}} + (\textbf{u} \cdot \nabla ) \textbf{u}
    = - \nabla p + \frac{1}{Re}\triangle \textbf{u} +
    N(\textbf{j}\times\textbf{B}),  \label{eq:NSE:momentum}
    \\
    \textbf{j}=-\nabla\phi + \textbf{u} \times \textbf{B},\,\,\,
    \nabla \cdot \textbf{u} = 0,\,\,\, \triangle \phi = \nabla
    \cdot(\textbf{u}\times\textbf{B}), \label{eq:NSE:Ohm}
    \end{eqnarray*} where $\bf{u}$ is velocity, $\bf{B}$ is external
magnetic field, $\bf{j}$ is electric current density, $p$ is
pressure, $\phi$ is electric potential. The above equations were
solved  by an own self-made 3D solver  for the same rectangular
duct as in the experiments. Technical numerical details are given
in \cite{Votyakov:Zienicke:FDMP:2006}.

The experimental data presented in
Fig.~\ref{Fig:Centerline},~\ref{Fig:2Dfields} were measured by one
of the authors, O.A., with the Ultrasonic Doppler Velocimeter
DOP2000, on the experimental setup described in details in
\cite{Andrejew:Kolesnikov:2006}. This setup is represented by the
Ga-In-Sn eutectic alloy flowing in the rectangular duct with sizes
(Length $\times$ Width $\times$ Height)=$2(L_x\times L_y\times
H)$=(50 $\times$ 10 $\times$ 2). The magnets assembled in a yoke
have horizontal sizes (Length $\times$ Width)=$2(M_x\times
M_y)$=(3 $\times$ 4) and distance between the poles $h=3$. All
numbers are in centimeters. Magnetic field strength measured in
the center of the magnetic gap is $B_0=0.4$T.

$Re$ and $N$ through the paper are defined with the parameters
$H$, $u_0$ and $B_0$ equal to half-height of the duct, the mean
flow velocity, and field intensity measured in the center of the
magnetic gap, correspondingly.

External magnetic field was modelled as a field from two permanent
magnets occupying a space $\Omega=\{|x|\leq M_x, |y|\leq M_y,
|z|>=h\}$, then ${\bf B(r)}=\nabla\int_{\Omega}{\bf B_d(r,r')}
d{\bf r'}$, where ${\bf B_d(r,r')}=\partial_{z'}(1/|{\bf r}-{\bf
r'}|)$ is a field created by a single magnetic dipole. After
algebraic calculations:
\begin{eqnarray*}
    B_{\gamma}(x,y,z)=\sum_{i,j,k=\pm 1} ijk\,F_{\gamma}(x-iM_x,
    y-jM_y,z-kh),
\end{eqnarray*} where $\gamma=x,y,z$ are for the magnetic field components,
$F_x(x,y,z)=\tanh^{-1}\frac{y}{r}$,
$F_y(x,y,z)=\tanh^{-1}\frac{x}{r}$,
$F_z(x,y,z)=\tan^{-1}\frac{zr}{xy}$, and $r=(x^2+y^2+z^2)^{-1/2}$.
Such an analytic approach is rather precise, as was tested by
comparing the calculated magnetic field with the field measured
experimentally.

\begin{figure*}[]
\begin{center}
  \subfigure[]{\label{Fig:DiagramConstraint}
    \includegraphics[width=8.0cm, clip]{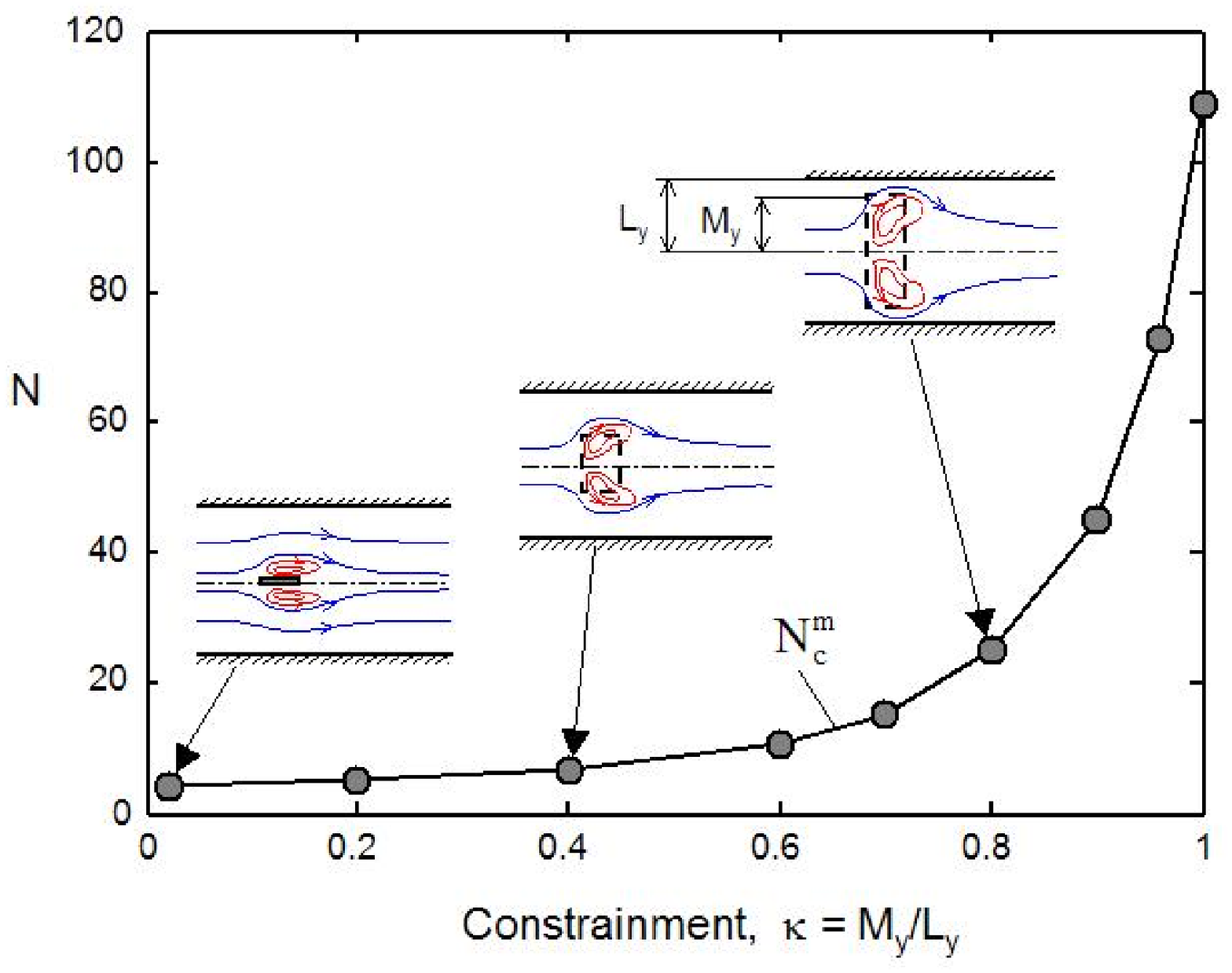}}
  \subfigure[]{\label{Fig:GlobalDiagram}
    \includegraphics[width=8.5cm, clip]{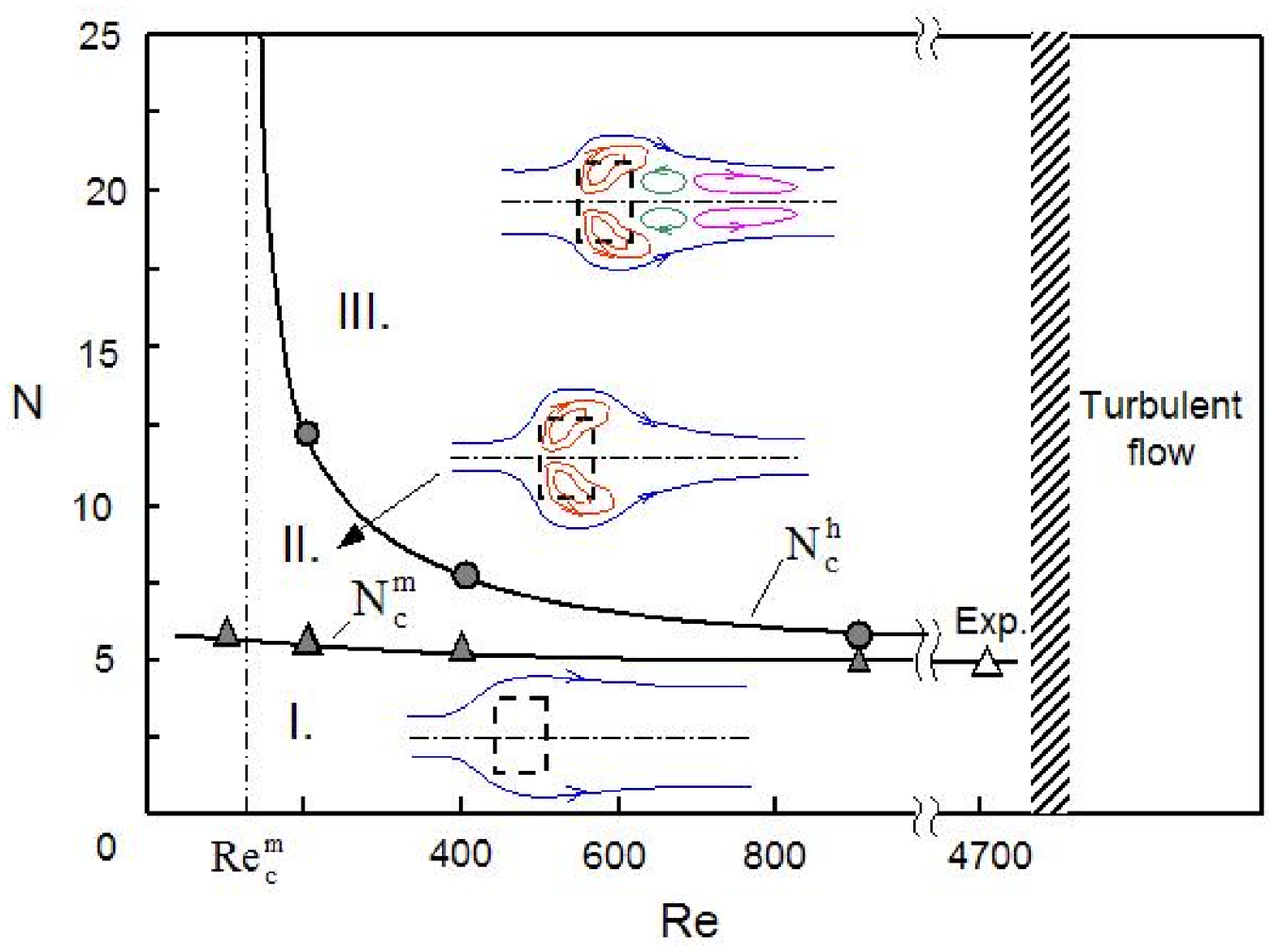}}
  \caption{{\label{Fig:Diagrams}(color online)}.{
  (a): Stability of inner magnetic vortices in dependence on
  constrainment $\kappa$ and interaction parameter $N$ at fixed $Re=100$.
  (b): Global stability diagram giving the existence regions of all three
  flow types described in this Letter in dependence  on ($Re$, $N$) at fixed $\kappa=0.4$.
    }
}
\end{center}
\end{figure*}


Let us now turn to final results presented as stability diagrams
in Fig.~\ref{Fig:Diagrams}. They summarize the flow behavior
around the magnetic obstacle dependent on the three parameters
$Re$, $N$ and $\kappa$, where $\kappa=M_y/L_y$ is the
constrainment ratio between the lateral size of the magnet,
$2M_y$, and the width of the channel, $2L_y$. The foresaid
simulated and experimental results correspond to a constrainment
ratio of $\kappa=0.4$.

In Fig.\ref{Fig:DiagramConstraint} the constrainment factor is
varied for the Reynolds number $Re=100$. The less is $\kappa$, the
less is the influence of side walls. The case of
$\kappa\rightarrow 0$ corresponds to a free flow. The larger is
$\kappa$, the more spanwise uniform is the braking Lorentz force.
Therefore, in order to induce inner vortices at larger $\kappa$ it
is necessary to apply  a larger critical interaction parameter
$N_c^m$ as one can see in Fig.~\ref{Fig:DiagramConstraint}. At
$\kappa \leq 0.5$ the critical value of $N_c^m$ is of the same
order of magnitude, $N_c^m\approx 6$, and after $\kappa\approx
0.8$ rapidly increases up to $N_c^m \approx 110$ at $\kappa=1$. We
believe that it is impossible to provoke magnetic vortices at high
$\kappa$ corresponding to spanwise uniform magnetic field.

At small constrainment ratio ($\kappa \ll 1$) and fixed $M_x$ the
magnetic obstacle becomes very slim (magnetic blade) compared to
the channel width. Then, attached vortices are not formed because
of well streamlined shape of the blade, while the magnetic
vortices are originated as before. Now, these magnetic vortices
are elongated along side walls of the magnetic blade and are
expanded in spanwise direction if interaction parameter $N$
increases.

%

Finally we present in Fig.~\ref{Fig:GlobalDiagram} the stability
diagram for $\kappa=0.4$. It shows, derived from a series of 3D
simulation and experiments, bifurcation lines $N_c^m(Re)$ and
$N_c^h(Re)$ separating the existence regions of flow without
vortices (I), with inner vortices (II), and flow including inner
magnetic vortices, connecting vortices and attached vortices
(III). To depict the borders between regions I-III we have
calculated several foothold points given in
Fig.~\ref{Fig:GlobalDiagram} by filled symbols. The open triangle
symbol at $Re=4700$ was obtained experimentally. Bifurcation
points were derived by monitoring the first (for inner) respective
second (for attached vortices) minimum on centerline curves as
shown in Fig.~\ref{Fig:Centerline}. These minima must be exactly
equal to zero at the bifurcation between two flow patterns. The
first bifurcation takes place slightly above $N\approx 5$ and is
almost independent of $Re$. If the magnetic vortices are present,
a minimum Reynolds number, $Re_c^m$, is necessary for the second
bifurcation to the appearance of attached vortices analogously  to
ordinary hydrodynamics.  Since we have never observed attached
vortices in the simulation with $Re=100$ and $N\leq 36$, $Re_c^m$
is given near 100. A series of experiments at
various $Re$ resulted in a critical value of $Re_c\approx 4700$
for inner vortices which corresponds to $N_c^m \approx 5$ fitting
well to the numerical values.

In summary, new phenomena are discovered for conducting liquid
flowing through a magnetic obstacle. One is the appearance of
stable inner vortices inside the obstacle at magnetic interaction
parameter $N \geq N_c^m$. The other is a new six-vortex flow
pattern appearing at $N \geq N_c^h(Re)$. This pattern is composed
of three pairs of vortices: inner, connecting, and attached. All
the vortices are found both in 3D numerical simulation and in
physical experiments.

\paragraph*{Acknowledgment} The authors express their gratitude
for financial support from the Deutsche Forschungsgemeinschaft
under grant ZI 667. Simulations were carried out at the John von
Neumann Institute (NIC) located in the Research Center J\"{u}lich.
We would like also to mention fruitful discussions with T.~Boeck,
C.~Karcher, S.~Smolentsev and S.~Cuevas.

\bibliography{./../../Bibtex/Channel_FDMP}

\begin{thebibliography}{14}
\expandafter\ifx\csname natexlab\endcsname\relax\def\natexlab#1{#1}\fi
\expandafter\ifx\csname bibnamefont\endcsname\relax
  \def\bibnamefont#1{#1}\fi
\expandafter\ifx\csname bibfnamefont\endcsname\relax
  \def\bibfnamefont#1{#1}\fi
\expandafter\ifx\csname citenamefont\endcsname\relax
  \def\citenamefont#1{#1}\fi
\expandafter\ifx\csname url\endcsname\relax
  \def\url#1{\texttt{#1}}\fi
\expandafter\ifx\csname urlprefix\endcsname\relax\def\urlprefix{URL }\fi
\providecommand{\bibinfo}[2]{#2}
\providecommand{\eprint}[2][]{\url{#2}}

\bibitem[{${\P}$}]{Votyakov}E-mail: evgeny.votyakov@tu-ilmenau.de

\bibitem[{\citenamefont{Feynman et~al.}(1964)\citenamefont{Feynman, Leighton,
  and Sands}}]{Feynman:Lectures:Vol2:1964}
\bibinfo{author}{\bibfnamefont{R.}~\bibnamefont{Feynman}},
  \bibinfo{author}{\bibfnamefont{R.}~\bibnamefont{Leighton}}, \bibnamefont{and}
  \bibinfo{author}{\bibfnamefont{M.}~\bibnamefont{Sands}},
  \emph{\bibinfo{title}{Lectures on physics. {V}ol II}}
  (\bibinfo{publisher}{{A}ddison-{W}esley}, \bibinfo{year}{1964}).

\bibitem[{\citenamefont{Shercliff}(1962)}]{Shercliff:book:1962}
\bibinfo{author}{\bibfnamefont{J.~A.} \bibnamefont{Shercliff}},
  \emph{\bibinfo{title}{The theory of electromagnetic flow-measurement}}
  (\bibinfo{publisher}{Cambridge University Press}, \bibinfo{year}{1962}).

\bibitem[{\citenamefont{Roberts}(1967)}]{Roberts:1967}
\bibinfo{author}{\bibfnamefont{P.~H.} \bibnamefont{Roberts}},
  \emph{\bibinfo{title}{An introduction to {M}agnetohydrodynamics}}
  (\bibinfo{publisher}{Longmans, Green}, \bibinfo{address}{New {Y}ork},
  \bibinfo{year}{1967}).

\bibitem[{\citenamefont{Davidson}(2001)}]{Davidson:book:2001}
\bibinfo{author}{\bibfnamefont{P.~A.} \bibnamefont{Davidson}},
  \emph{\bibinfo{title}{An introduction to {M}agnetohydrodynamics}}
  (\bibinfo{publisher}{Cambridge University Press}, \bibinfo{year}{2001}).

\bibitem[{\citenamefont{Moreau}(1990)}]{Moreau:book:1990}
\bibinfo{author}{\bibfnamefont{R.}~\bibnamefont{Moreau}},
  \emph{\bibinfo{title}{Magnetohydrodynamics}} (\bibinfo{publisher}{Kluwer},
  \bibinfo{year}{1990}).

\bibitem[{\citenamefont{Davidson}(1999)}]{Davidson:Review:1999}
\bibinfo{author}{\bibfnamefont{P.}~\bibnamefont{Davidson}},
  \bibinfo{journal}{Annual Review of Fluid Mechanics}
  \textbf{\bibinfo{volume}{31}}, \bibinfo{pages}{273} (\bibinfo{year}{1999}).

\bibitem[{\citenamefont{Thess et~al.}(2006)\citenamefont{Thess, Votyakov, and
  Kolesnikov}}]{Thess:Votyakov:Kolesnikov:2006}
\bibinfo{author}{\bibfnamefont{A.}~\bibnamefont{Thess}},
  \bibinfo{author}{\bibfnamefont{E.~V.} \bibnamefont{Votyakov}},
  \bibnamefont{and}
  \bibinfo{author}{\bibfnamefont{Y.}~\bibnamefont{Kolesnikov}},
  \bibinfo{journal}{Phys. Rev. Lett.} \textbf{\bibinfo{volume}{96}},
  \bibinfo{pages}{164501} (\bibinfo{year}{2006}).

\bibitem[{\citenamefont{Gelfgat et~al.}(1978)\citenamefont{Gelfgat, Peterson,
  and Shcherbinin}}]{Gelfgat:Peterson:Sherbinin:1978}
\bibinfo{author}{\bibfnamefont{Y.~M.} \bibnamefont{Gelfgat}},
  \bibinfo{author}{\bibfnamefont{D.~E.} \bibnamefont{Peterson}},
  \bibnamefont{and} \bibinfo{author}{\bibfnamefont{E.~V.}
  \bibnamefont{Shcherbinin}}, \bibinfo{journal}{Magnetohydrodynamics}
  \textbf{\bibinfo{volume}{14}}, \bibinfo{pages}{55} (\bibinfo{year}{1978}).

\bibitem[{\citenamefont{Gelfgat and
  Olshanskii}(1978)}]{Gelfgat:Olshanskii:1978}
\bibinfo{author}{\bibfnamefont{Y.~M.} \bibnamefont{Gelfgat}} \bibnamefont{and}
  \bibinfo{author}{\bibfnamefont{S.~V.} \bibnamefont{Olshanskii}},
  \bibinfo{journal}{Magnetohydrodynamics} \textbf{\bibinfo{volume}{14}},
  \bibinfo{pages}{151} (\bibinfo{year}{1978}).

\bibitem[{\citenamefont{Cuevas et~al.}(2006{\natexlab{a}})\citenamefont{Cuevas,
  Smolentsev, and Abdou}}]{Cuevas:Smolentsev:Abdou:2006}
\bibinfo{author}{\bibfnamefont{S.}~\bibnamefont{Cuevas}},
  \bibinfo{author}{\bibfnamefont{S.}~\bibnamefont{Smolentsev}},
  \bibnamefont{and} \bibinfo{author}{\bibfnamefont{M.}~\bibnamefont{Abdou}},
  \bibinfo{journal}{J. Fluid. Mech.} \textbf{\bibinfo{volume}{553}},
  \bibinfo{pages}{227 } (\bibinfo{year}{2006}{\natexlab{a}}).

\bibitem[{\citenamefont{Cuevas et~al.}(2006{\natexlab{b}})\citenamefont{Cuevas,
  Smolentsev, and Abdou}}]{Cuevas:Smolentsev:Abdou:PRE:2006}
\bibinfo{author}{\bibfnamefont{S.}~\bibnamefont{Cuevas}},
  \bibinfo{author}{\bibfnamefont{S.}~\bibnamefont{Smolentsev}},
  \bibnamefont{and} \bibinfo{author}{\bibfnamefont{M.}~\bibnamefont{Abdou}},
  \bibinfo{journal}{Phys. Rev. E} \textbf{\bibinfo{volume}{74}},
  \bibinfo{pages}{056301} (\bibinfo{year}{2006}{\natexlab{b}}).

\bibitem[{\citenamefont{Guckenheimer and
  Holmes}(1983)}]{Guckenheimer:Holmes:1983}
\bibinfo{author}{\bibfnamefont{J.}~\bibnamefont{Guckenheimer}}
  \bibnamefont{and} \bibinfo{author}{\bibfnamefont{P.}~\bibnamefont{Holmes}},
  \emph{\bibinfo{title}{Nonlinear Oscillations, Dynamical Systems and
  Bifurcations of Vector Fields}} (\bibinfo{publisher}{Springer Verlag},
  \bibinfo{address}{New York}, \bibinfo{year}{1983}).

\bibitem[{\citenamefont{Votyakov and
  Zienicke}(2007)}]{Votyakov:Zienicke:FDMP:2006}
\bibinfo{author}{\bibfnamefont{E.~V.} \bibnamefont{Votyakov}} \bibnamefont{and}
  \bibinfo{author}{\bibfnamefont{E.}~\bibnamefont{Zienicke}},
  \bibinfo{journal}{Fluid Dynamics and Materials Processing}
  \textbf{\bibinfo{volume}{3}}, \bibinfo{pages}{1} (\bibinfo{year}{2007}).

\bibitem[{\citenamefont{Andreev et~al.}(2006)\citenamefont{Andreev, Kolesnikov,
  and Thess}}]{Andrejew:Kolesnikov:2006}
\bibinfo{author}{\bibfnamefont{O.}~\bibnamefont{Andreev}},
  \bibinfo{author}{\bibfnamefont{Y.}~\bibnamefont{Kolesnikov}},
  \bibnamefont{and} \bibinfo{author}{\bibfnamefont{A.}~\bibnamefont{Thess}},
  \bibinfo{journal}{Phys. Fluids} \textbf{\bibinfo{volume}{18}},
  \bibinfo{pages}{065108} (\bibinfo{year}{2006}).

\end{thebibliography}


\end{document}